\documentclass[useAMS,usenatbib]{mn2e}
\usepackage{amsmath}
\usepackage{graphicx}
\usepackage{xcolor}
\numberwithin{equation}{section}

\newcommand{\be}{\begin{equation}}
\newcommand{\ee}{\end{equation}}
\newcommand{\ew}{e^{i\omega t}}
\newcommand{\gad}{\gamma_{\rm{ad}}}
\newcommand{\br}{\mathbf{r}}
\newcommand{\bv}{\mathbf{v}}
\newcommand{\om}{\omega^2}
\newcommand{\pz}{\cos(m\phi)\cos(kz)}
\newcommand{\pa}{\Phi_a}
\newcommand{\V}{\textbf{v}}
\newcommand{\X}{\textbf{x}}
\newcommand{\Wo}{W_1(f_1)}
\newcommand{\Wt}{W_2(f_1)}
\newcommand{\ddp}{\frac{d^2\Phi_0}{dr^2}}
\newcommand{\dpdr}{\frac{d\Phi_0}{dr}}
\newcommand{\dx}{d^3\textbf{x}}
\newcommand{\dv}{d^3\textbf{v}}
\newcommand{\fpe}{|f_0'(E)|}
\newcommand{\E}{\mathcal{E}}

\begin{document}
\title[Oscillations and stability of polytropic filaments]{Oscillations and stability of polytropic filaments}
\author[Patrick C. Breysse, Marc Kamionkowski, and Andrew Benson]{Patrick C. Breysse,$^{1}$\thanks{pbreysse@pha.jhu.edu (PCB); kamion@pha.jhu.edu (MK)} Marc Kamionkowski,$^{1}$\footnotemark[1] and Andrew Benson$^{2}$\thanks{abenson@obs.carnegiescience.edu}\\
$^{1}$Department of Physics and Astronomy, Johns Hopkins University, Baltimore, MD 21218 USA\\
$^{2}$Carnegie Observatories, Pasadena, CA 91101, USA}

\maketitle

\label{firstpage}

\begin{abstract}
We study the oscillations and stability of self-gravitating cylindrically symmetric fluid systems and collisionless systems.  This is done by studying small perturbations to the equilibrium system and finding the normal modes, using methods similar to those used in astroseismology.  We find that there is a single sequence of purely radial modes that become unstable if the adiabatic exponent is less than 1.  Nonradial modes can be divided into $p$ modes, which are stable and pressure-driven, and $g$ modes, which are are gravity driven.  The $g$ modes become unstable if the adiabatic exponent is greater than the polytrope index.  These modes are analogous to the modes of a spherical star, but their behavior is somewhat different because a cylindrical geometry has less symmetry than a spherical geometry.  This implies that perturbations are classified by a radial quantum number, an azimuthal quantum number, and wavelength in the $z$ direction, which can become arbitrarily large. We find that decreasing this wavelength increases the frequency of stable modes and increases the growth rate of unstable modes.   We use use variational arguments to demonstrate that filaments of collisionless matter with ergodic distribution functions are stable to purely radial perturbations, and that filaments with ergodic power-law distribution functions are stable to all perturbations.
\end{abstract}


\section{Introduction}
N-body simulations of the formation of large-scale structure reveal a rich web of filaments and voids, with clusters of galaxies forming at the intersections of filaments \citep{jenkins,wambs,ckc}.  Most of these simulations take into account only dark matter when carrying out their calculations, assuming that the baryonic matter will follow the perturbations to the dark matter.  However, simulations that include the separate evolution of baryons and dark matter find that the dynamics of the baryons have non-negligible impacts on the predicted structure \citep{ddk,ghh}.  A large fraction of the baryonic content of the Universe likely resides in this intergalactic medium \citep{cen}.  Observations of the Universe on the largest scales are now finding the filaments associated with this cosmic web \citep{turnetal,planck,bey}.

The behavior of these filamentary structures has important implications for the formation of structure in the universe.  In the prevailing cold-dark-matter paradigm, structure is formed in a bottom-up manner, while simulations of large-scale structure in warm-dark-matter models suggest that small-scale structure can arise from fragmentation of filaments in the cosmic web (Bode et al. 2001).  However, N-body simulations of filamentary structures often suffer from numerical artifacts introduced due to finite resolution \citep{hak,ww}.  Such simulations cannot model nearly the number of microscopic particles which would be present in an actual dark-matter distribution, and are restricted to a far smaller number of test particles.  This leads to two-body interactions between test particles that would not be present in a truly realistic scenario.  The result of these interactions tends to be an unphysical fragmentation of a filament along its axis of symmetry, which can create a larger amount of small scale power than would otherwise be present \citep{ww}.

We seek here to better understand the fragmentation of filamentary structures with an analytical study of the stability of self-gravitating fluid and collisionless systems.  We will start with an equilibrium model originally described by \citet{osta} of a cylindrically symmetric filament with a polytropic equation of state.  We then introduce linear-order perturbations to determine the normal modes of the filament, and classify these modes as stable or unstable.  This type of analysis is not feasible for filaments of collisionless material such as dark matter, but we still find that there are some general facts that can be determined about the stability of such systems.  Our results may not model fully the behavior of cosmic filaments, as our model is much simpler than real filaments produced by warm or cold dark matter models, but they provide a first step toward understanding this important problem analytically.

The relevance of this work is not limited to large-scale-structure formation.  Filamentary structures are common in the interstellar medium, and instabilities within these filaments likely create the dense cores in which stars form \citep{meyers}.  Tidal tails thrown off by merging galaxies can be modeled similarly \citep{sch,qc,cq}.  These tidal structures display clumping behavior which could be due to instabilities along their axes of symmetry.

The dynamics of cylindrically symmetric systems have seen some past study. The polytropic-equilibrium model used here was first derived by \citet{osta}.  \citet{sto} derived the magneto-hydrodynamical equilibrium of an isothermal filament, and \citet{elt}  showed that the same density profile applied to collisionless systems.  \citet{milgrom} showed that the work of \citet{elt} was valid in more general cases of systems which were neither isothermal nor axisymmetric, and generalized to theories of modified Newtonian dynamics.  Filaments consisting of both dark matter and baryonic matter were studied numerically by \citet{gt}, who found that the density of the baryonic fluid was well approximated by a power law with slope $-2$ in the central regions and -2.8 in the outer regions.

\citet{cf} studied the dynamics of a self-gravitating incompressible cylinder of uniform density, and \citet{ostb} generalized their work to a uniform-density compressible cylinder.  However, the stability of a fluid with realistically varying densities does not appear to have been studied. \citet{fp} describe in detail the behaviors of some types of cylindrical systems, but they are primarily interested in collisionless systems, and the distribution functions they consider appear to be fairly unusual.  \citet{qc} carried out a mostly analytic calculation of the dynamics of a tidal-tail structure based on some of the models of \citet{fp}. \citet{bessho} analyzed the behavior of a filament exposed to external radiation, and found that the radiation had a significant effect on the behavior of low density filaments.  Numerical simulations by \citet{kdgs} found that warm dark matter filaments tend to fragment into halos on the scale of the Jeans mass.

In this paper, we will start from an equilibrium model of a polytropic filament described in Section 2.  In section 3, we will examine the simplest normal modes of the system where all of the oscillation occurs in the radial direction, and section 4 will generalize this calculation to modes which oscillate in any direction.  Section 5 contains a discussion of the stability of collisionless filaments, and conclusions are made in Section 6.

\section{Equilibrium Configuration}

Following \citet{osta}, we define our equilibrium configuration by assuming the standard polytropic relation,
\be
P=K\rho^\gamma,
\ee
between pressure $P$ and density $\rho$ for some constants $K$ and $\gamma$ (see, for example, Kippenhahn and Weigert 1990, \S 19).  Equilibrium quantities are assumed to depend only on the distance from the center of the cylinder.   The equilibrium pressure, density, and potential $\Phi$ can be determined from the equation of hydrostatic equilibrium and Poisson's equation, which in cylindrical coordinates $(r,\ \phi\ z)$ can be written
\be
\frac{dP}{dr}=-\rho\frac{d\Phi}{dr},
\ee
and
\be
\frac{1}{r}\frac{d}{dr}\left(r\frac{d\Phi}{dr}\right)=4\pi G\rho.
\ee

Inserting equation (2.1) into equation (2.2) yields
\be
\frac{d\Phi}{dr}=-\gamma K\rho^{\gamma-2}\frac{d\rho}{dr}.
\ee
A relation between $\rho$ and $\Phi$ can be found by integrating equation (2.4).  The integration constant is chosen so that the potential is equal to 0 at the surface where $\rho=0$.  The resulting relation is 
\be 
\rho = C_n(-\Phi)^n,
\ee
where we have defined the quantities $n\equiv(\gamma-1)^{-1}$ and $C_n\equiv[K(n+1)]^{-n}$.

Substituting equation (2.5) into equation (2.3) gives a second-order differential equation 
\be
\frac{1}{r}\frac{d}{dr}\left(r\frac{d\Phi}{dr}\right)=4\pi GC_n(-\Phi)^n,
\ee
 for $\Phi$.  This can be written more simply by defining the dimensionless variables,
\be
\psi\equiv\frac{\Phi}{\Phi_c},\ \ \ \ s\equiv\frac{r}{b},
\ee
where $\Phi_c$ is the potential at $r=0$, and 
\be
b\equiv\left[4\pi GC_n(-\Phi_c)^{n-1}\right]^{-1/2}.
\ee
With these definitions, equation (2.6) becomes
\be
\frac{1}{s}\frac{d}{ds}\left(s\frac{d\psi}{ds}\right)=-\psi^n.
\ee
This is a modified form of the well-known Lane-Emden equation found for the potential profile of a spherical polytrope.  Since $\psi=\Phi/\Phi_c$, $\psi(s=0)$ must be equal to 1.  A second boundary condition can be chosen by requiring that the derivative of the potential be finite at the origin, which leads to the condition $d\psi/ds=0$ at the center of the cylinder.  

We now have a second-order equation and two boundary conditions which we can solve to get the potential profile.  Equation (2.9) has exact solutions when $n=0$, where $\psi(s) = 1-s^2/4$ and $n=1$, where $\psi(s) = J_0(s)$.  A third solution analogous to the $n=5$ solution of the spherical Lane-Emden equation may exist, but we have been unable to find it.  However, it is relatively easy to solve equation (2.9) numerically.  Figure 1 shows numerical solutions to equation (2.9) for $n=0$, 1, 3, and 5.    We assume that the total radius $R$ of the cylinder is known \emph{a priori}.  The maximum value of $s$ is the radius where $\psi=0$ so the value of $b$ can be found by taking the ratio of $R$ to this maximum $s$.  The value of $\Phi_c$ can then be found from equation (2.8), and density and pressure profiles can be found using equations (2.5) and (2.1).

\begin{figure}
\includegraphics[width=\columnwidth]{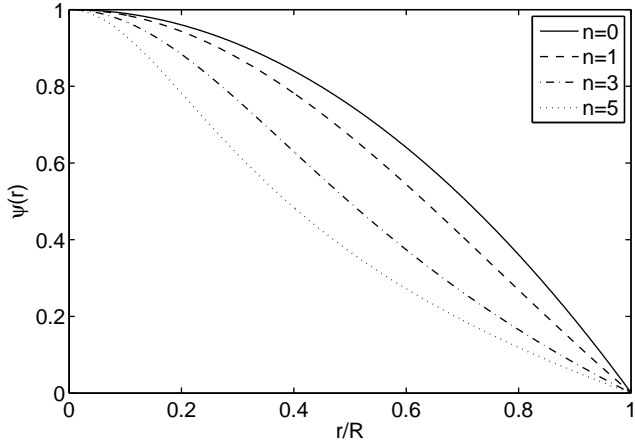}
\caption{Equilibrium potential as a function of radius $r$ in units of the maximum radius $R$ for polytropes with $n=0$,1, 3, and 5.  Higher values of $n$ correspond to softer equations of state, causing the potential profiles for higher $n$ values to be more centrally concentrated.}
\end{figure}

\section{Radial Oscillations}

We now seek to study the behavior of the system when it is perturbed away from equilibrium.  In this section we will consider purely radial perturbations, so the system retains its cylindrical symmetry.  We use a method similar to the one used in \S 38 of \citet{kw} adapted for a cylindrically symmetric system.  More details can be found in chapter 8 of \citet{cox}.  In subsection 3.1 we will determine the conditions under which we expect the system to be unstable, and in subsection 3.2 we will find the normal modes of the system and study their behavior.

\subsection{Stability Analysis}
When studying stellar oscillations it is common to use the mass enclosed within a given radius as the independent variable.  We adopt a similar convention here, but for cylinders we use the mass per unit length $\mu$.  Derivatives with respect to radius and $\mu$ are related by 
\be
\frac{d}{d\mu}=\frac{1}{2\pi r_0\rho_0}\frac{d}{dr_0},
\ee
where a subscript 0 denotes an equilibrium quantity.  We introduce a small perturbation with time dependence $e^{i\omega t}$ such that the pressure, radius, and density become 
\be
P(\mu,t) = P_0(\mu)(1+p(\mu)\ew), \nonumber
\ee
\be
r(\mu,t) = r_0(\mu)(1+x(\mu)\ew),
\ee
\be
\rho(\mu,t) = \rho_0(\mu)(1+d(\mu)\ew)\nonumber.
\ee
The perturbations are assumed to be small; i.e. $p/P_0$, $x/r_0$, and $d/\rho_0$ are all much less than one.  Perturbations with real values of $\omega$ are stable and oscillate around the equilibrium state; perturbations with imaginary values of $\omega$ are unstable and grow exponentially.

To find the values of $d$, $x$, and $p$, we need three independent equations.  We start by considering a thin shell of mass $d\mu$ per unit length a distance $r$ from the origin.  This shell experiences a force $f_P$ per unit area from the pressure gradient,
\be
f_P=-\frac{dP}{d\mu}d\mu,
\ee
and a force $f_g$ per unit area from the gravitational field,
\be
f_g=-\frac{G\mu}{\pi r^2}d\mu,
\ee
from the gravitational field.  Newton's second law then gives the equation of motion,
\be
\frac{1}{2\pi r}\frac{d^2r}{dt^2}=-\frac{dP}{d\mu}-\frac{G\mu}{\pi r^2},
\ee
for the shell.  We now substitute the perturbations from equation (3.2) into equation (3.3).  Since the perturbations are small, we neglect any term which is second order or higher in the quantities $d$, $x$, or $p$.  This yields
\be
\frac{d}{d\mu}\left(P_0p\right)=\left(2g_0+r_0\omega^2\right)\frac{x}{2\pi r_0},
\ee
where $g_0=2G\mu/r$, and we have cancelled the common factors of $\ew$.  Next we convert the mass derivative to a radial derivative using equation (3.1) to get 
\be
\frac{P_0}{\rho_0}\frac{dp}{dr_0}=\omega^2r_0x+g_0(p+2x).
\ee
Now consider the derivative of radius
\be
\frac{dr}{d\mu}=\frac{1}{2\pi r\rho},
\ee
with respect to $\mu$.  Substituting in the perturbations, linearizing, and replacing mass derivatives with radial derivatives as before yields
\be
r_0\frac{dx}{dr_0}=-2x-d.
\ee
In order to get a third independent equation, we make the additional simplifying assumption that the perturbations occur adiabatically.  The adiabatic approximation yields a relation,
\be
p=\gad d,
\ee
between the pressure and density perturbations, where the constant $\gad$ is the adiabatic exponent.  Note that $\gad$ is not necessarily the same as the equilibrium $\gamma$ found in equation (2.1).  Using equations (3.9) and (3.10) to remove the dependence on $p$ and $d$ from equation (3.7) yields a single second-order equation,
\be
x''+\left(\frac{3}{r_0}-\frac{\rho_0g_0}{P_0}\right)x'+\frac{\rho_0}{\gad P_0}\left(\omega^2+(2-2\gad)\frac{g_0}{r_0}\right)x=0,
\ee
for $x$, where primes indicate differentiation with respect to $r_0$

Multiplying equation (3.11) by $P_0r_0^3$ allows it to be written in the form of a standard Sturm-Liouville equation,
\be
\left(r_0^3P_0x'\right)'+\frac{r_0^3\rho_0}{\gad}\left(\omega^2+(2-2\gad)\frac{g_0}{r_0}\right)x=0,
\ee
 with eigenvalue $\omega^2$.  Solutions to equations of this form have several well-known properties.  There are an infinite number of eigenvalues $\omega_n$ with the property $\omega^2_{n+1}>\omega^2_n$.  Each eigenvalue $\omega^2_n$ has a corresponding eigenfunction $x_n$ with $n$ nodes in the interval $0<r_0<R_0$.  The lowest-order mode $x_0$ has no nodes and is called the fundamental.

These properties of the solutions to equation (3.12) allow us to determine the criteria for the stability of the system to radial perturbations.  We insert the fundamental eigenfunction $x_0$ into equation (3.12) and integrate from the center to the boundary.  The first term vanishes since $x'$ is required to be finite everywhere and $P_0(R_0)=0$, leaving
\be
\frac{\omega_0^2}{\gad}\int_0^{R_0}r_0^3\rho_0x_0dr_0+\frac{2-2\gad}{\gad}\int_0^{R_0}r_0^2\rho_0g_0x_0dr_0=0.
\ee
We can then solve this equation for the fundamental frequency,
\be
\omega_0^2=(2\gad-2)\frac{\int_0^{R_0}r_0^2\rho_0g_0x_0dr_0}{\int_0^{R_0}r_0^3\rho_0x_0dr_0}.
\ee
The fundamental is stable if $\omega_0^2$ is positive and unstable if it is negative.  Since $x_0$ has no nodes, it has the same sign everywhere in the cylinder, and both of the integrals in equation (3.14) have the same sign.  It is clear then that the fundamental is stable if $\gad>1$.  Since the fundamental has the lowest frequency, the system is stable to radial perturbations if $\gad>1$ and has at least one unstable mode if $\gad<1$.  This is similar to the criterion for radial perturbations of a spherical system, which are always stable if $\gad>4/3$ and unstable if $\gad<4/3$.  This result matches the one found in Section II of \citet{cf}.

\subsection{Normal Modes}

Now we seek to solve equation (3.12) and determine the behavior of the normal modes of the system.  To do this we replace the coefficients in equation (3.12) with functions of the polytrope quantities from Section 2.  Using equations (2.1), (2.5), (2.7), and (2.8) we can rewrite equation (3.12) as 
\begin{multline}
\frac{d^2x}{ds^2}+\left(\frac{3}{s}+\frac{n+1}{\psi}\frac{d\psi}{ds}\right)\frac{dx}{ds}-\left(\frac{b^2(n+1)}{\gad\Phi_c}\omega^2+\right. \\  \left.(2-2\gad)\frac{n+1}{\gad s}\frac{d\psi}{ds}\right)\frac{x}{\psi}=0.
\end{multline}
To solve this equation for $x$ we need boundary conditions.  The condition in the center is easy to see.  In order for $x$ to remain finite at $s=0$, we must have 
\be
\left(\frac{dx}{ds}\right)_{s=0}=0.
\ee
The boundary condition at the edge can be found by examining equation (3.7).  At the boundary of a polytrope, $P_0/\rho_0$ goes to 0, so the left hand side of equation (3.7) vanishes.  Thus, after replacing $p$ with $x$ and $dx/ds$, we have the condition 
\be
\left(\omega^2bS+g_0(2-2\gad)\right)x(S)-g_0\gad\left(\frac{dx}{ds}\right)_{s=S}=0.
\ee
Finally, since equation (3.15) is linear and homogeneous, we need a normalization condition.  This choice is entirely arbitrary, so we simply choose to set $x(R_0)=1$.  Some solutions to equation (3.15) for a system with $n=3$ and $\gad=4/3$ with these boundary conditions are shown in Figure 2.

\begin{figure}
\includegraphics[width=\columnwidth]{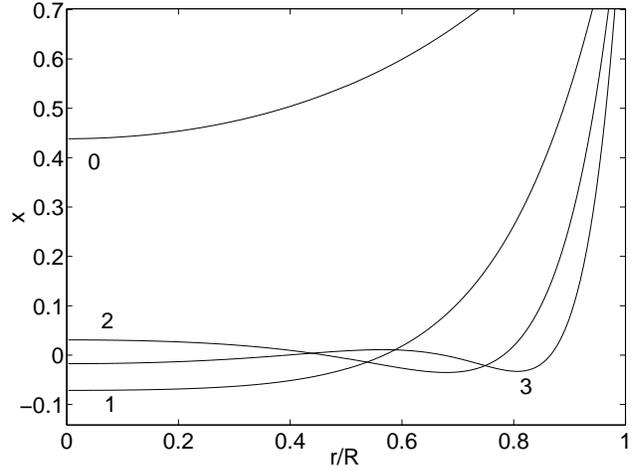}
\caption{First four radial eigenfunctions of an $n=3$ polytrope with $\gad=4/3$.  The fundamental is marked with a 0 and the first three higher-order modes are marked 1, 2, and 3.}
\end{figure}

\section{Nonradial Oscillations}
Now we move on to the more complicated case of nonradial perturbations, where material is free to move in any direction.  The calculations in this section use a similar method to the one \citet{hkt} uses to analyze adiabatic perturbations to spherical systems.  Additional information about spherical oscillations can be found in \citet{cox} and \citet{kw}.  We will follow the same path we did in section 3, first determining the criteria for instability then studying the normal modes of the system.

\subsection{Stability Analysis}

In the above case of radial perturbations we displaced a given shell of mass a small distance inward or outward from its initial position $r_0$.  Here we generalize this method, perturbing the system by moving every mass element a distance $\bxi(\br,t)$ from its initial position.  Consider here a single infinitesimal fluid element which has been moved to a new location without mixing with its surroundings.  If the perturbation is stable, this element will experience a force back towards its initial position, while if the perturbation is unstable, the element will feel a force directed away from where it came from.

There is a well-known quantity,
\be
\mathbf{A}\equiv\frac{1}{\rho_0}\nabla\rho_0-\frac{1}{\gad P_0}\nabla P_0,
\ee
from stellar-pulsation theory which determines whether or not this type of perturbation is stable \citep{cox}.  Since the equilibrium quantities in our model depend only on $r$, this simplifies to
\be
A=\frac{1}{\rho_0}\frac{d\rho_0}{dr}-\frac{1}{\gad P_0}\frac{dP_0}{dr}.
\ee
If $A>0$, the element feels a force away from its starting point and the system is convectively unstable (for a discussion of why, see section 17.2 of Cox 1980).  Thus we should expect to see unstable modes for any perturbation with $A>0$.  We can express this criterion as a condition on $\gad$ using the relation between $P_0$ and $\rho_0$ from equation (2.1).  Replacing $P_0$ in equation (4.2) yields
\be
A=\frac{1}{\rho_0}\frac{d\rho_0}{dr}-\frac{1}{\gad K\rho_0^\gamma}\frac{d}{dr}\left(K\rho_0^\gamma\right)
\ee
which simplifies to
\be
A=\frac{1}{\rho_0}\frac{d\rho_0}{dr}\left(1-\frac{\gamma}{\gad}\right).
\ee
Thus we expect unstable nonradial modes to exist when $\gad<\gamma$.

\subsection{Normal Modes}

We replace the equilibrium pressure, density, and potential with the respective perturbed quantities as in equations (3.2).  However, we must be a little more careful about how we define our perturbation.  A given perturbation can be described in two ways.  The Lagrangian description gives the perturbation to the mass element which was initially at position $\br$, and the Eulerian description gives the perturbation to the fluid at position $\br$.  For any quantity $q$, we will denote Lagrangian perturbations by 
\be
q=q_0+\delta q
\ee
and Eulerian perturbations by 
\be
q=q_0+q_1.
\ee
The two types of perturbations are related by 
\be
\delta q=q_1+\bxi\cdot\nabla q_0=q_1+\xi_r\frac{dq_0}{dr}
\ee
where $\xi_r$ is the radial component of the vector displacement $\bxi$ \citep{kw}.  The perturbed pressure, density, and potential are related by three equations: the Newtonian equation of motion,
\be
\rho\frac{\partial\bv}{\partial t}=-\nabla P-\rho\nabla\Phi,
\ee
the mass continuity equation,
\be
\frac{\partial\rho}{\partial t}+\nabla\cdot(\rho\bv)=0,
\ee
and Poisson's equation,
\be
\nabla^2\Phi=4\pi G\rho,
\ee
where $\bv$ in all of these equations is the fluid velocity.  The initial system is assumed to be stationary ($\bv_0=0$), so $\bv$ is just $\partial\bxi/\partial t$.  Inserting the Eulerian perturbations of $\rho$, $P$, and $\Phi$ into equations (4.8), (4.9), and (4.10) yields 
\be
\rho_0\frac{\partial^2\bxi}{\partial t^2}=-\rho_0\nabla\Phi_1-\rho_1\nabla\Phi_0-\nabla P_1,
\ee
\be
\frac{\rho_1}{\rho_0}=-\nabla\cdot\bxi-\frac{1}{\rho_0}\bxi\cdot\nabla\rho_0,
\ee
\be
\nabla^2\Phi_1=4\pi G\rho_1.
\ee
As before, we can relate perturbations of density and pressure by assuming the perturbations are adiabatic.  The Lagrangian density and pressure perturbations are thus related by 
\be
\frac{\delta P}{P_0}=\gad\frac{\delta\rho}{\rho_0}.
\ee

As before we assume that the time dependence of the perturbation is $\ew$.  With this time dependence, equation (4.11) becomes
\be
\om\bxi=-\frac{\rho_1}{\rho_0^2}\frac{dP_0}{dr}\hat{r}+\nabla\Phi_1+\frac{1}{\rho_0}\nabla P_1,
\ee
where $\hat{r}$ is the unit vector in the radial direction, and we have used equation (2.2) to replace the derivative of $\Phi_0$ with the derivative of $P_0$.

Equation (4.15) is a vector equation, so we can consider each of its components individually.  We start with the radial component:
\be
\om\xi_r=-\frac{\rho_1}{\rho_0^2}\frac{dP_0}{dr}+\frac{\partial\Phi_1}{\partial r}+\frac{1}{\rho_0}\frac{\partial P_1}{\partial r}.
\ee
To simplify the next few steps of the calculation, we rearrange the derivatives in equation (4.16) so that the radial derivative acting on $P_1$ acts on $P_1/\rho_0$ instead:
\be
\om\xi_r=\frac{P_1}{\rho_0^2}\frac{d\rho_0}{dr}+\frac{\partial}{\partial r}\left(\frac{P_1}{\rho_0}\right)-\frac{\rho_1}{\rho_0^2}\frac{dP_0}{dr}+\frac{\partial\Phi_1}{\partial r}.
\ee
Next we use equation (4.7) to express the mass continuity equation (4.12) in terms of $\delta\rho$:
\be
\frac{\delta\rho}{\rho_0}=-\nabla\cdot\bxi.
\ee
Now we replace $P_1$ in the first term and $\rho_1$ in the second term by $\delta P$ and $\delta\rho$ and use equations (4.14) and (4.18) to replace these Lagrangian perturbations with $\nabla\cdot\bxi$, which results in
\be
\om\xi_r=\frac{\partial}{\partial r}\left(\frac{P_1}{\rho_0}\right)-Av_s^2\nabla\cdot\bxi+\frac{\partial\Phi_1}{\partial r},
\ee
where 
\be
v_s^2=\frac{\gad P_0}{\rho_0}
\ee
is the sound speed and $A$ is from equation (4.2).  

The angular and longitudinal parts of equation (4.15) can be written 
\be
\omega^2\xi_\phi=\frac{1}{r\rho_0}\frac{\partial P_1}{\partial\phi}+\frac{1}{r}\frac{\partial\Phi_1}{\partial\phi}
\ee
and 
\be
\omega^2\xi_z=\frac{1}{\rho_0}\frac{\partial P_1}{\partial z}+\frac{\partial\Phi_1}{\partial z}.
\ee
With these, we can now write the nonradial components of $\bxi$ in terms of $P_1$:
\be
\bxi(\br)=\xi_r\hat{r}+\frac{\partial}{\partial\phi}\left(\frac{P_1}{r\om\rho_0}+\frac{\Phi_1}{r\om}\right)\hat{\phi}+\frac{\partial}{\partial z}\left(\frac{P_1}{\om\rho_0}+\frac{\Phi_1}{\om}\right)\hat{z}.
\ee
To proceed, we need to carry out a separation of variables on the three components of $\bxi$ and on $P_1$ and $\Phi_1$.  For spherical stars, the angular dependence of these quantities is usually expanded in spherical harmonics $Y_{lm}(\theta,\phi)$.  For our cylindrical system, we express the nonradial dependence with the function $\pz$ where $m$ is a dimensionless integer and $k$ is the wave number of the perturbation along the $z$-axis.  With this separation, $P_1$ and $\Phi_1$ become
\be
P_1(\br)=P_a(r)\pz,
\ee
and
\be
\Phi_1(\br)=\Phi_a(r)\pz,
\ee
where the radial dependences are in the unknown functions $P_a(r)$ and $\Phi_a(r)$.  The displacement $\bxi$ becomes
\begin{multline}
\bxi(\br)=\left[\xi_a(r)\hat{r}+\left(\frac{P_a(r)}{r\om\rho_0}+\frac{\Phi_a}{r\om}\right)\hat{\phi}\frac{\partial}{\partial\phi}+\right. \\ \left.\left(\frac{P_a(r)}{\om\rho_0}+\frac{\Phi_a}{\om}\right)\hat{z}\frac{\partial}{\partial z}\right]\pz.
\end{multline}
We can simplify this further by defining the quantity,
\be
\xi_t\equiv\frac{P_a}{r\om\rho_0}+\frac{\Phi_a}{r\om},
\ee
which when substituted into equation (4.26) yields
\be
\bxi(\br)=\left[\xi_a(r)\hat{r}+\xi_t\hat{\phi}\frac{\partial}{\partial\phi}+r\xi_t\hat{z}\frac{\partial}{\partial z}\right]\pz.
\ee

Now we have to express equations (4.18) and (4.19) in terms of our unknowns $\xi_a$ and $\xi_t$.  To do this, we first need to calculate the divergence of $\bxi$, which in cylindrical coordinates is
\be
\nabla\cdot\bxi=\left[\frac{1}{r}\frac{d}{dr}(r\xi_a)-\left(\frac{m^2}{r}+rk^2\right)\xi_t\right]\pz.
\ee
The right hand side of equation (4.18) is thus known, and we can rewrite the left hand side using hydrostatic equilibrium as well as equations (4.7) and (4.14):
\be
\frac{\delta\rho}{\rho_0}=\frac{1}{v_s^2}\left(\om r\xi_t-\Phi_a-g_0\xi_a\right),
\ee
where $g_0=\nabla\Phi_0$ is the equilibrium gravitational field.  The mass continuity equation therefore yields
\be
r\frac{d\xi_a}{dr}=\left(\frac{g_0r}{v_s^2}-1\right)\xi_a+\left(m^2+r^2k^2-\frac{\omega^2r^2}{v_s^2}\right)\xi_t+\frac{r}{v_s^2}\pa.
\ee
In equation (4.19), the divergence of $\bxi$ can be replaced by equation (4.29), and the $P_1$ term can be replaced with equation (4.27), leaving 
\be
r\frac{d\xi_t}{dr}=\left(\frac{Ag_0}{\om}+1\right)\xi_a-\left(1+Ar\right)\xi_t+\frac{A}{\om}\pa.
\ee

We now have three unknowns ($\xi_t$, $\xi_r$, and $\pa$) and two equations.  We require one more, which we can get from equation (4.10).  We can replace the density perturbation with a pressure perturbation using equations (4.7) and (4.14):
\be
\rho_1=\frac{1}{v_s^2}\left(P_1+\xi_r\frac{dP_0}{dr}\right)-\xi_r\frac{d\rho_0}{dr}.
\ee
Next we separate out the nonradial dependence as before and use equation (4.27) to express the density perturbation in terms of our unknowns,
\be
\rho_a=-A\rho_0\xi_a+\frac{r\om\rho_0}{v_s^2}\xi_t-\frac{\rho_0}{v_s^2}\Phi_a,
\ee
where as with the other quantities we have defined $\rho_1=\rho_a(r)\pz$.  Expanding the Lagrangian on the left-hand side of equation (4.10) then yields
\begin{multline}
\frac{d^2\pa}{dr^2}+\frac{1}{r}\frac{d\pa}{dr}-\left(\frac{m^2}{r^2}+k^2\right)\pa= \\ 4\pi G\left(-A\rho_0\xi_a+\frac{r\om\rho_0}{v_s^2}\xi_t-\frac{\rho_0}{v_s^2}\Phi_a\right).
\end{multline}

Equations (4.31), (4.32), and (4.35) can be further simplified by introducing the Brunt-V\"{a}is\"{a}l\"{a} frequency,
\be
N^2\equiv-Ag_0=-g_0\left(\frac{1}{\rho_0}\frac{d\rho_0}{dr}-\frac{1}{\gad P_0}\frac{dP_0}{dr}\right),
\ee
the Lamb frequency,
\be
S_l^2\equiv\left(\frac{m^2}{r^2}+k^2\right)v_s^2,
\ee
and the transverse wavenumber
\be
k_t^2\equiv\frac{m^2}{r^2}+k^2=\frac{S_l^2}{v_s^2}.
\ee
We also introduce the perturbation 
\be
g_a\equiv d\pa/dr.
\ee
to the gravitational field.  With these quantities, we obtain
\be
r\frac{d\xi_a}{dr}=\left(\frac{k_t^2g_0r}{S_l^2}-1\right)\xi_a+r^2k_t^2\left(1-\frac{\om}{S_l^2}\right)\xi_t+\frac{rk_t^2}{S_l^2}\pa
\ee
\be
r\frac{d\xi_t}{dr}=\left(1-\frac{N^2}{\om}\right)\xi_a+\left(\frac{rN^2}{g_0}-1\right)\xi_t-\frac{N^2}{g_0\om}\pa
\ee
\begin{multline}
r\frac{dg_a}{dr}=\frac{4\pi GrN^2\rho_0}{g_0}\xi_a+\frac{4\pi Gr^2\om\rho_0k_t^2}{S_l^2}\xi_t+ \\ rk_t^2\left(1-\frac{4\pi G\rho_0}{S_l^2}\right)\pa-g_a.
\end{multline}

We now have in equations (4.39--42) a system of equations with eigenfunctions $\xi_a$, $\xi_t$, $\pa$, and $g_a$ and eigenvalue $\om$.

In order to solve this system we need to set appropriate boundary conditions.  First examine the ratio $\delta P/P_0$ at the surface of the system, which is related to our unknowns by 
\be
\frac{\delta P}{P_0}=\frac{\rho_0}{P_0}\left(\omega^2r\xi_t-\pa-g_0\xi_a\right).
\ee
The ratio on the left hand side should remain finite just below the surface, but the ratio $\rho_0/P_0$ diverges at the surface.  In order for the right hand side to remain finite, we must set the quantity in parenthesis equal to zero.  This means that at the maximum radius $R$, $\xi_a$ and $\xi_t$ are related by 
\be
\omega^2R\xi_t(R)-\pa(R)-g_0(R)\xi_a(R)=0
\ee

A second boundary condition can be found from the fact that the potential must match up with a solution of Laplace's equation at $r=R$.  For $r\geq R$, we choose the solution 
\be
\pa(r\geq R)=CJ_m(kR)
\ee
where $C$ is a normalization constant and $J_m$ is a Bessel function of the first kind.  Taking the derivative of equation (4.45) and using the Bessel function derivative identity yields the condition 
\be
g_a(R)-\frac{k\pa(R)}{J_m(kR)}\left(J_{m-1}(kR)-J_{m+1}(kR)\right)=0.
\ee
This condition is valid in all but two cases.  The derivative of the $m=0$ Bessel function is $-J_1(kr)$, so the boundary condition for $m=0$ modes is
\be
g_a(R)+\frac{k\pa(R)J_1(kR)}{J_0(kR)}=0.
\ee
When $k=0$, the second term of equation (4.46) diverges.  However, when $k=0$ we can choose the form,
\be
\pa(r\geq R)=Cr^m,
\ee
for the potential, yielding the condition,
\be
g_a(R)-\frac{m}{R}\pa(R)=0.
\ee

One central boundary condition arises from the requirement that the derivative of the potential be finite everywhere, which forces
\be
g_a(0)=0.
\ee
The other central boundary condition can be found by making the approximation that close to the center,
\be
\xi_a = r^a\sum_\nu X_\nu r^\nu,
\ee
and
\be
\xi_t = r^b\sum_\nu Y_\nu r^\nu.
\ee
Near the center, the equilibrium quantities $A$ and $g_0$ approach 0, so equations (4.40) and (4.41) become approximately
\be
r\frac{d\xi_a}{dr}\approx-\xi_a+m^2\xi_t
\ee
\be
r\frac{d\xi_t}{dr}\approx\xi_a-\xi_t.
\ee
Substituting in the lowest order ($\nu=0$) terms from equations (4.39) and (4.40) yields
\be
r^aX_0=-r^aX_0+m^2r^bY_0,
\ee
and
\be
r^bY_0=r^aX_0-r^bY_0.
\ee
Since $X_0$ and $Y_0$ are dimensionless, we see from dimensional analysis that $a$ must be equal to $b$.  We then have
\be
(a+1)X_0=m^2Y_0
\ee
and
\be
(a+1)Y_0=X_0.
\ee
From equations (4.45) and (4.46) it is clear that $a=m+1$ and
\be 
X_0=mY_0
\ee
which in turn means that the boundary condition at $r=0$ is 
\be
\xi_a(0)=m\xi_t(0).
\ee

Finally, as with the radial case we have a linear, homogeneous system, so we need a normalization condition.  We choose for simplicity to set $\xi_a(R)=1$.

We can now solve equations (4.39-42) numerically with boundary conditions (4.38) and (4.48).    Since the eigenvalue $\om$ appears nonlinearly in these equations, the normal mode spectrum is not the simple Sturm-Liouville case of the radial modes.  Nonradial modes can be divided into two types, called $p$ modes and $g$ modes by analogy to similar modes of spherical stars. The restoring force for $p$ modes is primarily pressure, and the restoring force for $g$ modes is primarily gravity.  

Each mode is given a number which for most modes corresponds to the number of nodes in $\xi_a$.  The exception is the $g1$ mode, which often has no nodes.  For a given $m$ and $k$, the $g1$ mode is the highest frequency g-mode, and the p1-mode is the lowest frequency p-mode.  There is also a single mode known as the f-mode with no nodes in $\xi_a$ with a frequency between that of the g1 and p1 modes. The $p$ modes are always stable, but the stability of the $g$ modes depends on the value of $\gad$.  As predicted by the stability criterion from section 3.2, the $g$ modes are stable if $\gad>\gamma$ and unstable if $\gad<\gamma$. If $\gad=\gamma$, $N^2=0$ and the $1/\omega^2$ term vanishes from equation (4.41).  The system thus reduces to an ordinary Sturm-Liouville problem and the $p$ modes are the only modes present.

Figures 3 and 4 show the eigenfunctions for the first three p-modes and g-modes of a system with $n=3$ and $\gad=1.6$, and Figure 5 shows $\xi_a$ and $\xi_t$ for the f-mode of the same system.  All modes in figures 3, 4, and 5 are calculated for $m=1$ and $k=0$.  Note that $p$ modes tend to act more in the outer regions of the cylinder, while $g$ modes tend to act closer to the center.  Also, in g-modes, the first node of $\xi_t$ occurs before the first mode of $\xi_a$, while the reverse is true for p-modes.

\begin{figure}
\includegraphics[width=\columnwidth]{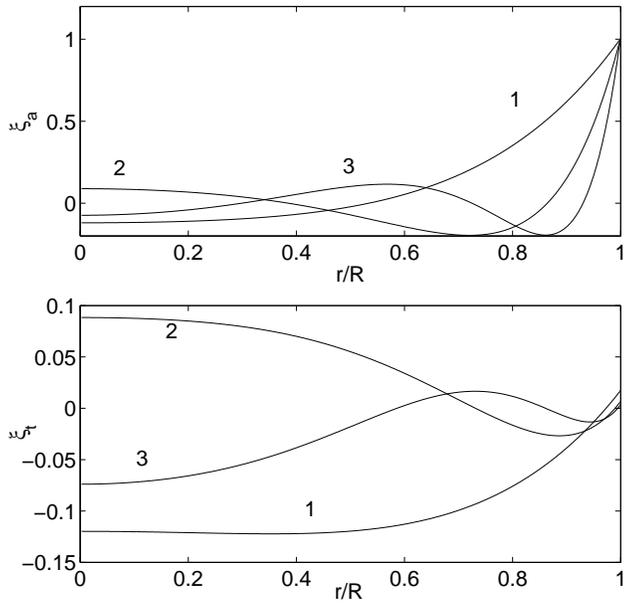}
\caption{Eigenfunctions of the first three $p$ modes of a system with $n=3$ and $\gad=1.6$.  Modes are calculated for $m=1$ and $k=0$.  The top panel shows the $\xi_a$ eigenfunction and the bottom panel shows the $\xi_t$ eigenfunction.}
\end{figure}

\begin{figure}
\includegraphics[width=\columnwidth]{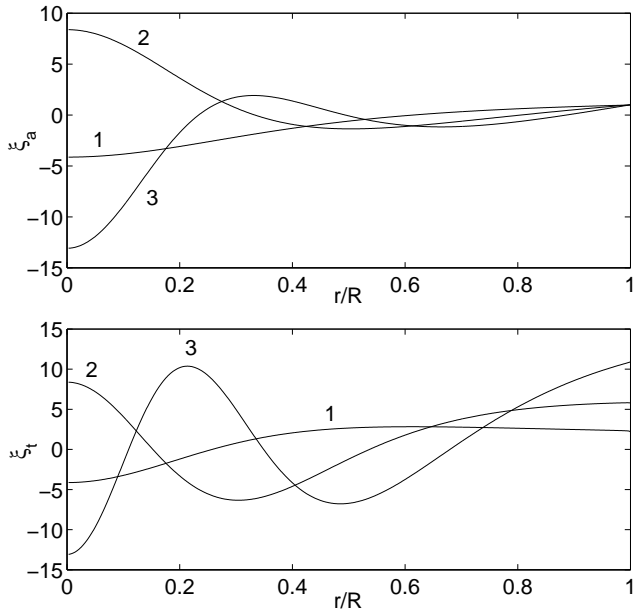}
\caption{Eigenfunctions of the first three $g$ modes of the system from Figure 3. The top panel shows the $\xi_a$ eigenfunction and the bottom panel shows the $\xi_t$ eigenfunction.}
\end{figure}

\begin{figure}
\includegraphics[width=\columnwidth]{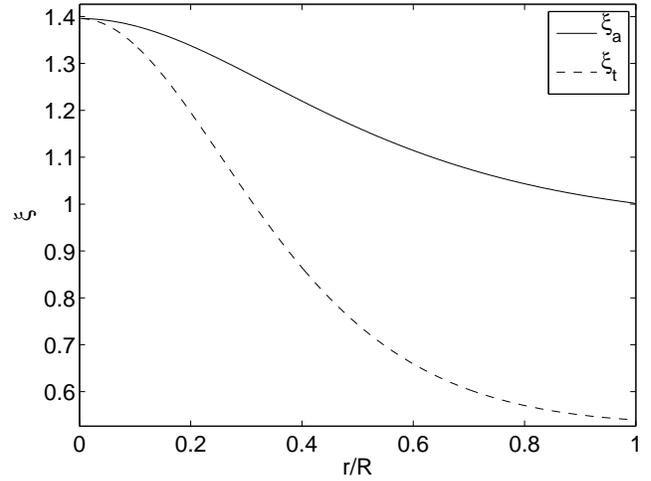}
\caption{Eigenfunctions of the $f$ mode of the system from Figure 3.}
\end{figure}

Figure 6 shows the eigenvalues $\om$ for different $m$'s.  The marked points are the eigenvalues.  Since $m$ has to be an integer, the lines between the points have no physical meaning and only serve to connect frequencies of the the same mode.  Figures 7 and 8 explore the behavior of p and $g$ modes for different values of $k$.  Figure 9 does the same for the unstable $g$ modes of a system with $\gad<\gamma$.  From figures 6--9 it is clear that allowing the fluid to move tangentially makes stable modes more stable and unstable modes more unstable.  This is similar to the result Ostriker (1964b) obtained for a homogeneous cylinder.

\begin{figure}
\includegraphics[width=\columnwidth]{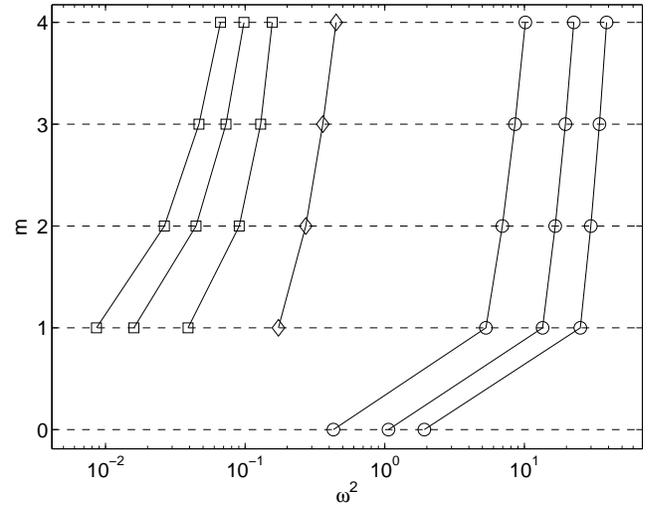}
\caption{Eigenvalues for $p$, $g$, and $f$ modes at different values of $m$ for the system in Figure 3 with $k=0$. Circles are frequencies of $p$ modes; squares are frequencies of $g$ modes; and diamonds are frequencies of the $f$ mode.  The lines have no physical meaning; they only serve to connect frequencies of the same mode.}
\end{figure}

\begin{figure}
\includegraphics[width=\columnwidth]{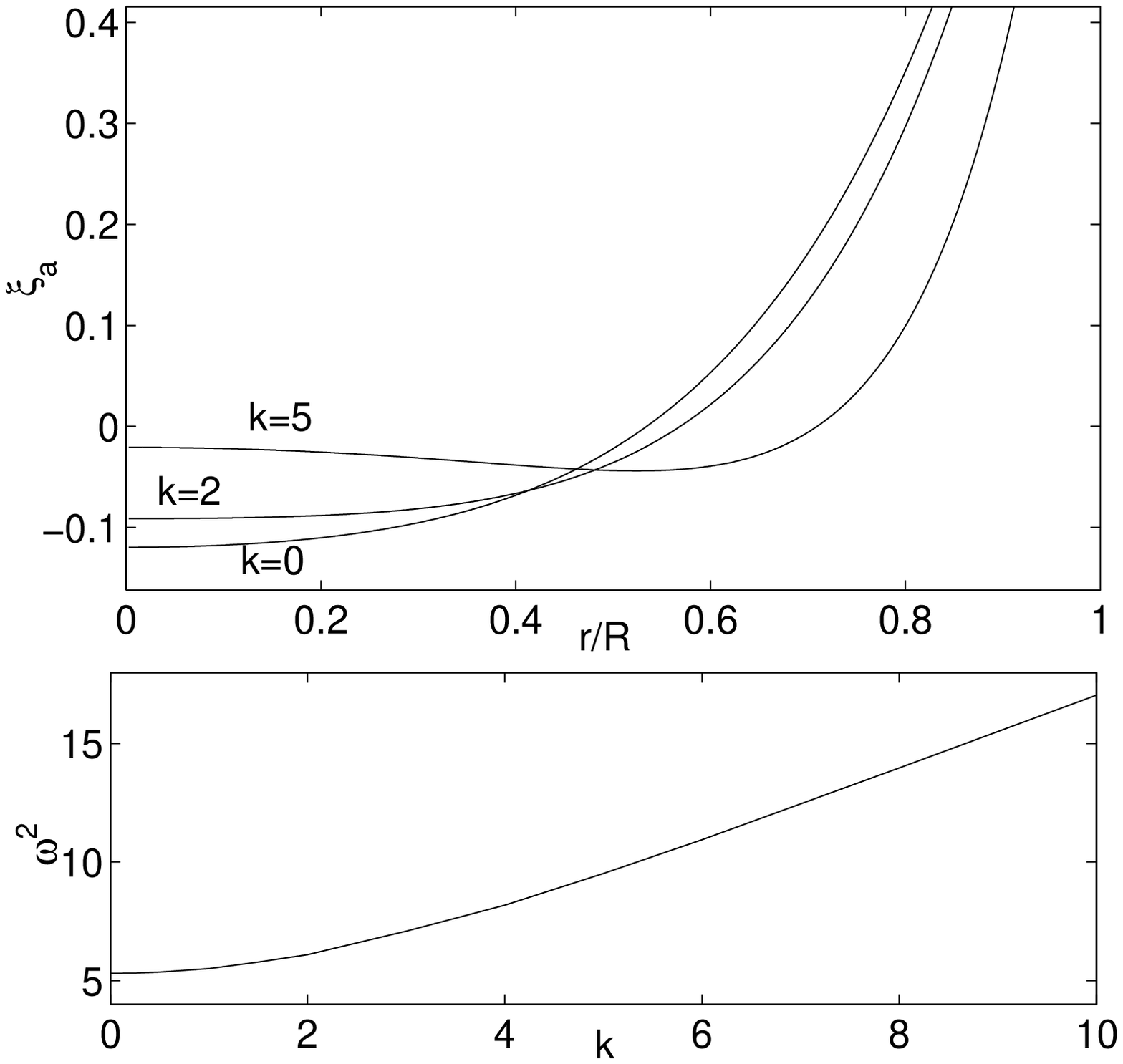}
\caption{Behavior of $p$ modes with increasing $k$ for the system from Figure 3.  Modes are calculated for $m=1$.  The top panel shows how the $\xi_a$ eigenfunction changes with $k$, and the bottom panel shows how $\om$ changes with $k$.}
\end{figure}

\begin{figure}
\includegraphics[width=\columnwidth]{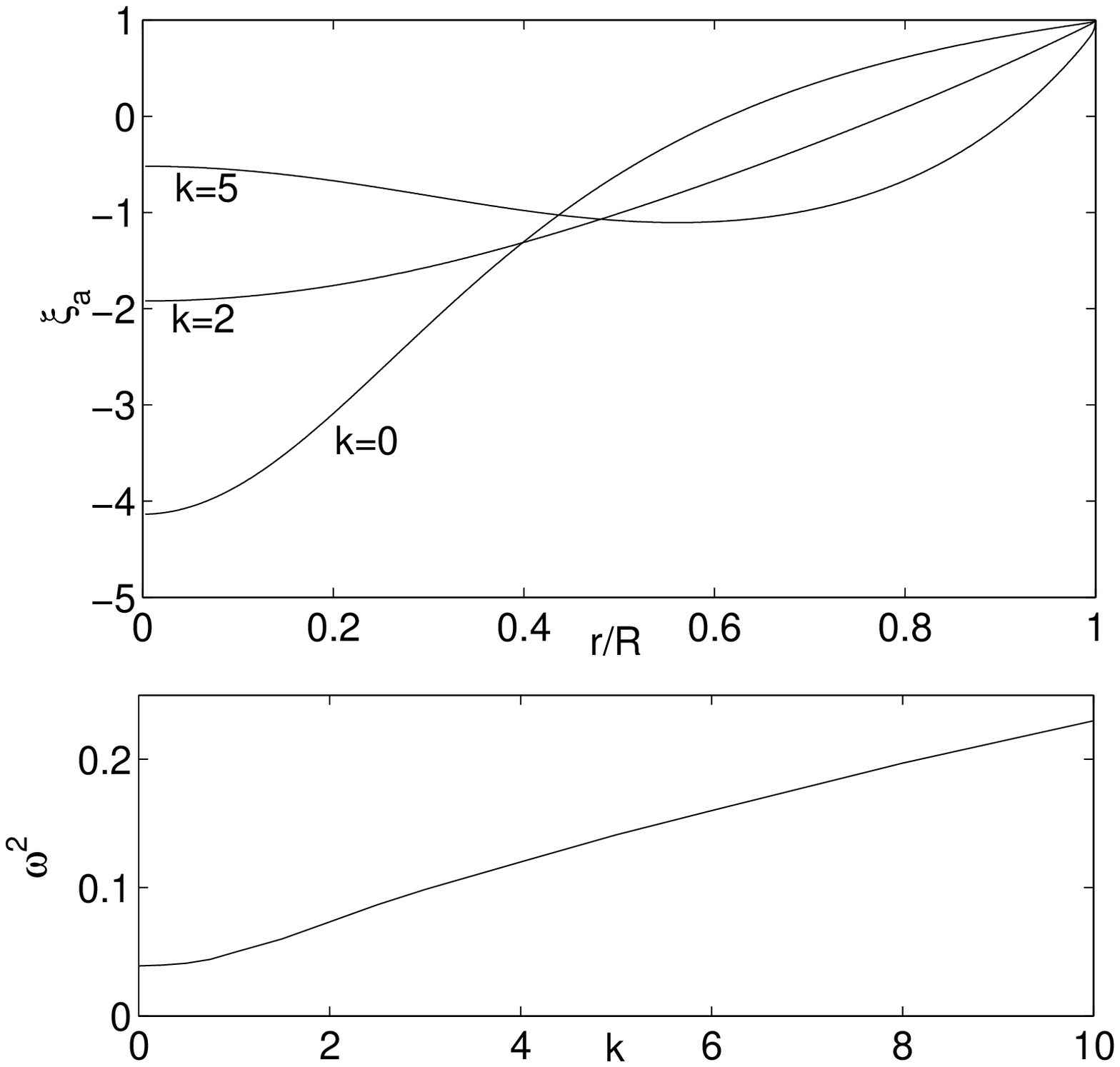}
\caption{Behavior of $g$ modes with increasing $k$ for the system from Figure 3.  Modes are calculated for $m=1$.  The top panel shows how the $\xi_a$ eigenfunction changes with $k$, and the bottom panel shows how $\om$ changes with $k$.}
\end{figure}

\begin{figure}
\includegraphics[width=\columnwidth]{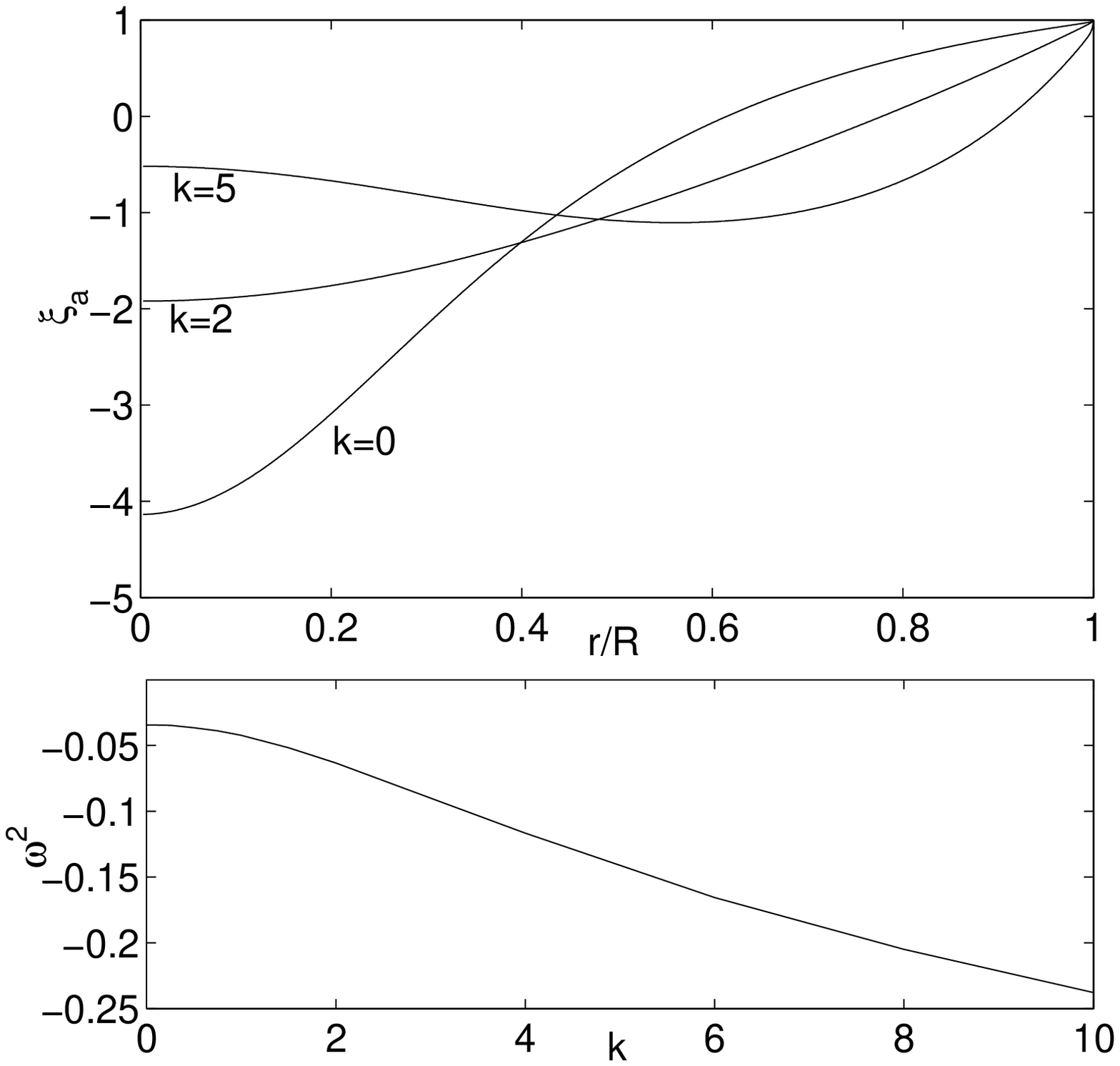}
\caption{Behavior of unstable $g$ modes with increasing $k$ for system with $\gad<\gamma$.  Modes are calculated for $m=1$.  The top panel shows how the $\xi_a$ eigenfunction changes with $k$, and the bottom panel shows how $\om$ changes with $k$.}
\end{figure}

\section{Stability of Collisionless Filaments}
The collisionless nature of dark matter filaments makes the kind of approach carried out in Sections 3 and 4 prohibitively difficult.  Such filaments are not pressure supported, so all of the calculations would need to be done in 6-dimensional phase space rather than the 3-dimensional space required by the fluid filaments.  We will therefore not attempt to calculate the normal modes of a collisionless filament in this paper.  However, it is still possible to draw conclusions about the stability of dark matter filaments even if the details of their oscillations are not known.  Most of the calculations in this section are based on the energy principle from section 5.4 of \citet{BT}, which essentially states that any perturbation which increases the total energy of a system is stable and any which decreases the total energy is unstable.

\subsection{Radial Stability}
The Doremus-Feix-Baumann Theorem states that a spherically symmetric collisionless system with an ergodic distribution function ($f(\X,\V)=f(E)$) is stable to radial perturbations \citep{BT}.  We can prove that the same holds true for cylindrically symmetric systems.  This proof is based on the variational principle for collisionless systems presented in Binney and Tremaine chapter 5, which in two dimensions states that a collisionless system with equilibrium distribution function $f_0(\X,\V)$ is stable to a perturbation $f(\X,\V)=f_0(\X,\V)+f_1(\X,\V)$ if the quantity,
\begin{multline}
W(f_1)\equiv\int\frac{f_1^2(\X,\V)}{|f_0'(E)|}d^2\X d^2\V- \\ G\int\frac{f_1(\X,\V)f_1(\X'\V')}{|\X-\X'|}d^2\X d^2\V d^2\X'd^2\V'\geq0
\end{multline}
where $f_0'$ is the derivative of $f_0$ with respect to $E$.  For a self gravitating system, $f_0'$ is assumed to be negative.

We start by introducing polar coordinates in velocity space $(v,\phi_v)$.  The volume element in these coordinates is 
\be
d^2\V=vdvd\phi_v.
\ee
The energy in these coordinates is $E=v^2/2+\Phi_0(r)$ where $\Phi_0$ is the unperturbed potential.   The angular momentum is $L = rv\mathrm{sin}\phi_v$.  Changing the volume element into coordinates $(E,L)$ yields 
\be
d^2\V=\frac{1}{y}dEdL
\ee
where $y=(2r^2(E-\Phi_0(r))-L^2)^{1/2}$.

Consider the second term $\Wt$ of the variational principle.  This term is equal to twice the gravitational potential energy of the perturbation, so it can be rewritten as
\be
\Wt=-\frac{1}{G}\int r\left(\frac{d\Phi_1}{dr}\right)^2dr,
\ee
where $\Phi_1$ is the potential of the perturbation.  From Gauss's Law, 
\be
\frac{d\Phi_1}{dr}=\frac{2G\mu(r)}{r}=\frac{4\pi G}{r}\int_0^r\int f_1(r',\V')d^2\V'dr'
\ee
\be
=\frac{4\pi G}{r}\int_0^r\int f_1(r',\V')\frac{1}{y}dEdLr'dr'.
\ee
Next we remove $f_1$ from this equation by defining the quantity $g(r,E,L)$ through
\be
f_1(r,\V)\equiv\frac{f_0'(E)y(r,E,L)}{r}\frac{d}{dr}(y(r,E,L)g(r,E,L)).
\ee
Expressed in terms of $g$, the derivative in equation (19) becomes
\be
\frac{d\Phi_1}{dr}=\frac{f\pi G}{r}\int_0^r\int f_0'(E)\frac{d}{dr}(yg)dEdLdr,
\ee
and $\Wt$ becomes 
\be
\Wt=-8\pi^2G\int\frac{1}{r}\left(\int f_0'ygdEdL\right)^2dr,
\ee

Now consider the first term $\Wo$ of the variational principle.  Written in terms of $g$, $dE$, and $dL$, this term is 
\be
\Wo=-2\pi\int\frac{1}{r}f_0'y\left(\frac{d}{dr}(yg)\right)^2dEdLdr.
\ee
In order to determine the sign of the total $W(f_1)$ we will use Schwarz's inequality, which states that for two functions $A(x)$ and $B(x)$ 
\be
\int A^2dx\int B^2dx\geq\left(\int ABdx\right)^2.
\ee
For this proof we choose $A=(-f_0'y)^{1/2}$ and $B=Ag$.  Then, Schwarz's inequality in terms of these two functions is 
\be
\int f_0'ydEdL\int f_0'yg^2dEdL\geq\left(\int f_0'ygdEdL\right)^2.
\ee
The term on the right-hand side of this inequality appears in $\Wt$, so we can say that 
\be
\Wt\geq-8\pi G\int\frac{1}{r}\int f_0'ydEdL\int f_0'yg^2dEdLdr.
\ee  

The integral of $f_0'y$ in equation (26) can be integrated by parts with respect to $E$, resulting in 
\be
\int f_0'ydEdL=-\int f_0\frac{dy}{dE}dE_1dL_1.
\ee
The boundary terms vanish because $y=0$ at the minimum energy and $f_0=0$ at the maximum energy.  Converting equation (26) into an integral over $d^2\V$ and integrating yields
\be
\int f_0'ydEdL=-r^2\int f_0d^2\V=-r^2\rho_0(r),
\ee
where $\rho_0$ is the density of the unperturbed system.

With these results, the full variational principle can be simplified to 
\be
W(f_1)\geq-2\pi\int\frac{f_0'y}{r}\left(\left[\frac{d}{dr}(yg)\right]^2-4\pi r^2G\rho_0g^2\right)dEdLdr.
\ee
The term in square brackets can be integrated by parts, and the boundary terms will vanish because $y=0$ at $r=0$ and as $r$ goes to infinity.  The resulting integral has several terms, one of which contains $d^2g/dr^2$.  Integrating this term by parts a second time yields
\begin{multline}
W(f_1)\geq-2\pi\int\frac{f_0'y}{r}\left[\left(y\frac{dg}{dr}\right)^2- \right. \\ \left. g^2\left(\left(\frac{dy}{dr}\right)^2+y\frac{d^2y}{dr^2}-\frac{y}{r}\frac{dy}{dr}-4\pi G\rho_0r^2\right)\right]dEdLdr.
\end{multline}

This relation can be further simplified using Poisson's Equation 
\be
\nabla^2\Phi=\ddp+\frac{1}{r}\dpdr=4\pi G\rho_0.
\ee
Using the definition of $y$ we can write 
\be
\frac{1}{r}\frac{d}{dr}\left(r\frac{dy^2}{dr}\right)=-2\ddp-\frac{6}{r}\dpdr.
\ee
Combining equations (31) and (32) gives 
\be
y\frac{d^2y}{dr^2}+\left(\frac{dy}{dr}\right)^2-\frac{y}{r}\frac{dy}{dr}+4\pi G\rho_0r^2=-2r\dpdr.
\ee
With this, equation (30) simplifies to 
\be
W(f_1)\geq-2\pi\int\frac{f_0'y}{r}\left(\left(y\frac{dg}{dr}\right)^2+2rg^2\dpdr\right)dEdLdr.
\ee
Since $f_0'(E)$ is always negative and $d\Phi_0/dr$ is always positive in a realistic system, the quantity on the right hand side is always positive and so is $W(f_1)$.  Thus the system is stable to purely radial perturbations.

The basic physical mechanism behind this stability has to do with the behavior of the constituent particles of the system when it is either compressed or expanded \citep{BT1}.  Compressing the filament radially releases gravitational potential energy, which is transferred into the translational motion of the particles in the system.  This increased velocity resists the compression and pushes the system back towards equilibrium.  The opposite happens if the filament is expanded.  What happens to the perturbed filament depends on the relative size of the change in gravitational potential energy compared to the change in kinetic energy.  The above proof demonstrates that the equilibrium configuration is a minimum energy state, so the change in kinetic energy is always greater, and radial displacements evolve back towards equilibrium.

\subsection{Nonradial Stability}
We have no easy way to determine the stability of a filament with arbitrary distribution function.  However, using Antonov's First Law \citep{BT} we can draw conclusions about filaments with a specific class of distribution functions.  Antonov's First Law states that a collisionless system with an ergodic distribution function and $f_0'(E)<0$ is stable if a fluid system with the same density distribution and with $\gamma=\gad$ is stable.   This theorem can be fairly easily proven using equation (5.1) and Chandrasekhar's variational principle, which states that a fluid system is stable to density perturbation $\rho_1$ if 
\be
W_f(\rho_1)=\int\left|\frac{d\Phi}{d\rho}\right|_0\rho_1^2{\X}\dx-G\int\frac{\rho_1(\X)\rho_1(\X')}{|\X-\X'|}\dx\dx'
\ee
is greater than or equal to 0.  This is a necessary and sufficient condition for stability, so we can also state that if a fluid system is stable, then $W_f\geq0$.

Consider an equilibrium fluid system with density $\rho_0$ and a collisionless system with distribution function $f_0$ such that $\rho_0=\int f\dx\dv$.  Now introduce a perturbation $\rho_1=\int f_1\dx\dv$ to both systems.  Using equation (5.11) with $A=|f_0'(E)|$ and $B=f_1/|f_0'(E)|^{1/2}$ gives
\be
\int\frac{f_1^2}{\fpe}\dv\geq\frac{\left(\int f_1\dv\right)^2}{\int\fpe\dv}.
\ee
We then integrate both sides with respect to $\X$ and add the second term of equation (5.1) to both sides.  The result simplifies to
\be
W(f_1)\geq W_f(\rho_1),
\ee
where $W(f_1)$ comes from equation (5.1), and $W_f(\rho_1)$ comes from equation (5.22).  Thus, since a stable fluid system with $\gamma=\gad$ will have $W_f\geq0$, we must also have $W\geq0$ and the collisionless system is stable.

From section 4 we know that a polytropic fluid filament with $\gamma=\gad$ is stable, so we can apply Antonov's First Law if we can find a distribution function that yields the same density distribution.  Consider an ergodic power-law distribution function of the form,
\be
f(\E)=F(\E)^{n-1},
\ee
where $F$ is an arbitrary constant and $\E\equiv-E=-v^2/2-\Phi$.  The density profile of this filament can be related to the potential $\Psi\equiv-\Phi$ by integrating this distribution function over velocity,
\be
\rho(r)=2\pi\int f(\E)dv=2\pi F\int_0^\Psi\E^{n-1}d\E
\ee
\be
\rho(r)=\frac{2\pi F}{n}\Psi^n(r)=c_n(-\Phi)^n.
\ee
This relation is the same as the one obtained for the polytropic fluid in equation (2.5).  The potential can then be obtained from Poisson's equation, which turns out exactly as it does for the fluid in equation (2.9).  Thus, dark matter filaments with distribution functions given by equation (5.25) have the same density profiles as fluid filaments with polytrope index $n$, and by Antonov's first law, such filaments are stable.

It is important to note that Antonov's first law is only a sufficient condition for stability and not a necessary one.  If a fluid system is unstable and $W_f$ in equation (5.24) is negative it does not require the corresponding collisionless system to be unstable.

\section{Conclusions}

Structures with cylindrical symmetry arise throughout astrophysics, from star-forming regions, to tidal tails, to the cosmic web, and instabilities in these structures may have implications for star and galaxy formation.  We have undertaken here an analytic study of the stability of models of polytropic fluid filaments.  We find that instabilities in fluid filaments are convective in nature, and are enhanced by tangential fluid motion.  We also find that, despite the qualitative differences between spherical and cylindrical geometry, the basic characteristics of the normal modes of a cylinder are fairly similar to those of a sphere.  Both geometries have a single series of radial modes which are all stable if $\gad$ is greater than some critical value.  This critical value is $4/3$ for a sphere, and we confirm the result of \citet{cf} that the critical value is 1 for a cylinder.  

Non-radial modes for both geometries can be divided into pressure-driven $p$ modes and gravity-driven $g$ modes with a single $f$ mode in between.  For both systems, the $p$ modes are always stable, and the $g$ modes can be either stable or unstable depending on the value of $\gad$. The criterion for instability for these modes is the Schwarzschild criterion for convective instability.  The modes of a cylindrical system also have dependence on the wavenumber $k$ due to the fact that the cylindrical geometry is less symmetric than the spherical geometry.  We find that increasing $k$ increases the oscillation frequency of stable modes and increases the growth rate of unstable modes.    This is in agreement with the results of \citet{ostb}, who found that the instabilities in a homogenous filament are convective in nature and are enhanced by nonradial oscillations.

Though the above detailed mode analysis could not be applied to collisionless filaments, the results for fluid filaments made it possible to draw some conclusions about the stability of certain types of collisionless cylinders using variational arguments.  We showed that filaments with ergodic distribution functions are stable to purely radial perturbations.  Also, because they have the same density profile as polytropic fluid filaments, we were able to demonstrate that collisionless filaments with ergodic power-law distribution functions are stable to all perturbations.  This is in contrast to the fragmentation seen in simulations performed by \citet{kdgs}.  However, the filaments seen to fragment in these simulations were found in a larger N-body simulation, and may not satisfy the conditions necessary for our analysis to apply, i.e. that the filament be well approximated by an infinitely long, self-gravitating, cylindrically symmetric collisionless cylinder with a specific form of distribution function.  Thus we do not necessarily expect the results of \citet{kdgs} to exactly match ours.

To reconcile our analytic results with those of numerical simulations, it would therefore be useful to directly measure the distribution function of filaments in simulations of both cold and warm dark matter. If those distribution functions can be characterized by simple, analytic forms, the methodology applied in this work could be used to address the stability of these cosmological filaments. This would then provide guidance as to whether physical fragmentation is ever to be expected, or if the fragmentation seen will always be due to numerical artifacts, no matter how much the resolution of simulations is increased.

The models studied here for both fluid and collisionless systems are fairly simple.  Realistic fluid systems would have additional complications such as turbulence and magnetic fields, and collisionless systems can easily have far more complicated distribution functions.  Other work on similar systems shows how different models can give significantly different results.  For example, \citet{qc} find that a homogeneous collisionless filament develop instabilities at critical wavelengths similar to the Jean's instability, and \citet{cf} find that a magnetic field can eliminate instabilities in an incompressible filament.  The methods used here could fairly easily be adapted to less simplified situations.  For example, Antonov's first law could be used to double check that instabilities found in dark matter simulations are real and not artifacts by examining a fluid model with the same density profile.  

\section*{acknowledgments}
This work was supported by DoE SC-0008108 and NASA NNX12AE86G.

\end{document}